*A self-assembled periodic nanoporous framework in aqueous solutions of the DNA tetramer GCCG*

Gregory P. Smith[a], Tommaso P. Fraccia[b,c], Chenhui Zhu[d], Tommaso Bellini[b,2], Noel A. Clark[a,2]

[a]Department of Physics and Soft Materials Research Center, University of Colorado, Boulder, CO, 80309-0390

[b]Dipartimento di Biotecnologie Mediche e Medicina Traslazionale, Università degli Studi di Milano, via Fratelli Cervi 93, I-20090 Segrate (MI), Italy

[c]Institut Pierre-Gilles de Gennes, Chimie Biologie et Innovation, ESPCI Paris, PSL University, CNRS, 6 rue Jean Calvin, 75005, Paris, France

[d]Advanced Light Source, Lawrence Berkeley National Laboratory, Berkeley, CA 94720 USA

[2]*To whom correspondence may be addressed: noel.clark@colorado.edu*

***Abstract***

We broaden the realm of DNA nanoscience by demonstrating that single-component aqueous solutions of ultrashort DNA oligomers can spontaneously organize into 3D mesoscale frameworks. The DNA tetramer 5′-GCCG-3′ forms B-type double helices by Watson-Crick (WC) pairing into tiled brickwork-like duplex chains, which arrange into nematic and columnar liquid crystal phases. At concentrations $c_{\mathrm{DNA}} > 400$mg/mL, these solutions facilitate the slow nucleation and growth of a body-centered cubic lattice, which maintains its low density (DNA volume fraction ~0.2) up to solution concentrations of $c_{\mathrm{DNA}} \sim 700$ mg/ml. This lattice is an array of valence-8 nodes, each formed by a pair of quadruplexes of GCCG terminal guanosines, connected by 6-molecule long WC duplex struts. Atomistic simulations confirm the stability of such arrangements.

*Introduction*

The key role of DNA in biology stems from the selective interactions between the side groups (bases) on pairs of polymeric NA chains direct molecular self-assembly, enabling matching sequences of large numbers of bases, $N$, to package and transmit genetic information in duplex structures of paired helical strands. These structures are stabilized in solutions of arbitrarily small concentration, $c$, by the selective Watson-Crick (WC) adenosine-thymine and guanine-cytosine base-pairing motif [1], and by the columnar stacking in the duplex of the aromatic hydrocarbon nanosheets formed by the paired bases [2,3]. Nucleobases can also form and exploit non-WC bonds, among which are quadruplex strands of stacked guanosine tetramers (G4s), relevant for example in the functioning of telomers [4].

A more recent development has been the advent of DNA nanoscience, which has enabled the creation of an amazing variety of active and passive functional nanoscale structures via DNA oligomers self-assembled by similar base-pairing motifs. Appropriately designed combinations of NA strand lengths and base sequences can orchestrate programmed hierarchical self-assembly in solution on length scales ranging from nanometer to micrometer [5,6]. These biological and nanoscience application of NAs are generally based on self-assembly pathways which are operative at low concentration in aqueous solution, designed such that the specific intermolecular interactions are strong enough to bind single oliogomers at very low osmotic pressure of the nucleic acid solutes.

We have pursued in recent years a broad study of nucleic acids in the quite different regime where $N$ is small, ($1 \leq N \lesssim 10$), and $c_{\mathbf{DNA}}$ is large, in the range 200 mg/mL $\lesssim c_{\mathbf{DNA}} \lesssim$ 1200 mg/mL [7,8,9,10,11,12,13,14,15,16,17]. At these concentrations, which are up to ~2/3 of that of dry DNA ($c_{\mathrm{DNA}}$ ~ 1700 mg/mL) [18, 16], nucleic acids self-assemble into three dimensional collective structures and phases that depend on $N$, $c_{\mathrm{DNA}}$, temperature, and the base sequences of the NA oligomers. Generally, as $N$ is decreased the NA osmotic pressures and therefore the concentrations required for collective self-assembly increase. In such solutions, where the oligomers are complementary or self-complementary, WC-like local organization and double helix formation can be obtained for any $N$, *i.e.* even $N = 1$, as in solutions of the monomer nucleotides [16]. The principally observed 3D self-assembly motifs that result are the packing of these duplex base-stacked columns into parallel arrays to form bulk nematic liquid crystal (LC) phases, or, at higher concentration, spontaneous arrangement of the columns on a hexagonal lattice to form columnar LCs, similar to that found in long DNA [19,20] at similar concentrations [7-16]. These phases are identifiable by x-ray diffraction and microscopic study of their birefringent optical textures [7,16].

This observation of WC duplex organization, in absence of the phospho-diester polymerization that stabilizes the double helix at low concentration, provides evidence for the hypothesis motivating the above-mentioned work, namely that LC spatial organization of single bases at high concentration could be a templating and selection mechanism for the appearance of double helical and polymeric nucleic acids in the approach to the RNA world in early life [14,15]. However, we have found that the small-*N*, regime also proves to be a rich area for the general exploration of novel soft-matter collective behaviors in NAs at higher concentrations. Here we report a nucleic acid system, obtained in single component aqueous solutions of the DNA tetramer 5′-GCCG-3′, in which DNA can spontaneously assemble into a low-density 3D mesoscale framework of DNA rods, extending the horizon of nucleic acid nanofabrication to include components comprising a single molecular species, and ultra-short NAs.

The molecule investigated here is the DNA oligomer 5′-GCCG-3′ in which the phosphodiester four-base chain is -OH terminated on both its 5′ and 3′ ends. We will refer to this molecule simply as GCCG. Synthesis, purification details are presented in Materials and Methods. Of the work cited above the most relevant to the results to be presented here is that of Ref. [13], our prior study of the four-base DNA oligomer 5′-GCCGp-3′ (GCCGp ), in which the phosphodiester four-base chain is -OH terminated on its 5′ end, and -$PO_4$ terminated on its 3′ end. In aqueous solution GCCGp presented the phase sequence [isotropic (ISO) – nematic (NEM) – columnar (COL)] with increasing concentration, a scenario similar to that found in solutions of longer $N < 20$ oligomers. However, in addition, GCCGp exhibited a second (reentrant) isotropic (ISO2) phase at the highest $c_{\mathrm{DNA}}$ values, an observation not precedented in our earlier studies of hundreds of other oligomers. This ISO2 phase was tentatively proposed in Ref. [13] to be due to the formation of random networks of oligomer chains, driven by association of the terminal G's favored at high $c_{\mathrm{DNA}}$.

The work presented here on GCCG was motivated by this observation, and by a basic theme emerging from our extensive exploration of N < 20 oligomers, namely that as *N* decreases their collective behavior becomes more and more sensitive to terminal composition. We report here that, while GCCG and GCCGp exhibit similar overall phase behavior, certain conditions enable GCCG to form a new "BCCX" mesoscopic framework phase that we analyze and understand to be a spontaneous cooperative ordering of a combination of WC and G4 molecular bonds. In order to facilitate GCCG – GCCGp comparison, GCCGp phase behavior is included with that of GCCG in ***Fig. 1*** and in the *Supplementary Information* as ***Figs. S1*** and ***S2***.

The LC phases in these systems can be understand in terms of the WC "brickwork" duplex chaining mode of GCCGp and GCCG, sketched in ***Fig. 2A.*** and ***Fig. S1***. In this chaining there is no distinction between duplex formation and linear aggregation. Rather, hybridization and chaining necessarily take place in a single continuous process, in which the bonding of each GCCG molecule to an existing chain (***Fig. S1B***) and the merging of different chains (***Fig. S1C***) are characterized by a free energy gain. This condition leads to equilibrium states characterized by a broad polydisperse distribution of lengths, ranging from single dissociated oligomers to multi-molecular aggregates with a characteristic length depending on interaction strength and temperature, $T$ as generally expected in isodesmic self-assembly [21]. Rods which are poly-disperse in length have a strong tendency to form nematic and columnar LC phases, the dominant LC phases observed here and in a large variety of $N < 20$ NA oligomer solutions [22].

***Results***

Exploration of the phase behavior of solutions vs. solution concentration $c_{DNA}$ and $T$ by depolarized transmission optical microscopy (DTOM) and x-ray scattering enabled division of the ($c_{DNA}$,$T$) plane into regions exhibiting distinct isotropic and birefringent optical domains characteristic of textures made by DNA LC phases, including the isotropic (ISO), nematic (NEM), helical nematic (NEM*) and fluid columnar (COL) [7,13]. The phase behavior was assessed by preparing solutions, of chosen DNA concentrations in the range 100 mg/mL < $c_{DNA}$ < 800 mg/mL, and filling them into planar glass cells or thin-wall x-ray capillaries. These solutions were made with or without added salt, the latter case making the corresponding native Na+ counterion concentration in the aqueous partition 0.27M $\lesssim c_{Na+} \lesssim$ 3.9M in the above $c_{DNA}$ range. Such fixed concentration samples were then temperature cycled at rates sufficiently slow to enable consequent formation of domains and concentration equilibration of the nematic and columnar phases. It is typical of DNA LCs that the transitions among these phases are first order. In addition, they are strongly hysteretic at the high concentrations being studied here (~5% to ~45% of neat DNA), conditions which combine to make regions of two- and three- phase coexistence quite common in the optical and x-ray cells. ***Fig. 1B*** shows the phase diagram of CGGC obtained from the optical and x-ray experiments, along with that of GCCGp in ***Fig. 1A***, from Ref.[13], for comparison.

GCCGp and GCCG both exhibit the ISO phase of weakly aggregated oligomers at lower $c_{DNA}$ and higher $T$, with the NEM and then the COL phase the first LC phases appearing with increasing concentration at lower $T$. In the GCCGp diagram of ***Fig. 1A*** the single phase regions are solid-colored, while two-phase coexistence areas are indicated in white. In the GCCG diagram of ***Fig. 1B***, most of the diagram area exhibits two or three phases in coexistence. In this circumstance the concentrations of the coexisting phases ($c_{phase1,}$ $c_{phase2,}$ etc.) will generally be different from each other and from $c_{DNA}$, which sets the overall average. In both phase diagrams LC ordering is quite stable, found at higher temperatures and lower concentrations than in some longer blunt-ended oligomers (6-10mers) [7,8,11]. This is a result of the absence of the weaker A-T hydrogen bonding, the WC duplexes being stabilized exclusively by the stronger G-C bonding

Remarkably, in both CGGC and GCCGp the low concentration ISO phase range extends to the highest concentrations, a feature not found previously in any other short-DNA oligomer solutions, which typically form birefringent biaxial columnar liquid crystals of columnar crystals, at high $c_{\mathrm{DNA}}$ [7,8,11]. In GCCGp the ISO transitions subtly into a second isotropic phase, the ISO2p, which persists at high $c_{\mathrm{DNA}}$ upon cooling to room temperature. This high-$c_{\mathrm{DNA}}$ isotropic was attributed to the enhanced tendency for G-G-G-G tetra-base aggregation, promoted at high DNA concentration relative to CG base pairing by mass action, resulting in the formation of an isotropic network gel of duplex strands cross-linked by G quadruplexes [13]. GCCG exhibits a quite similar phase scenario, clearly transitioning to a second isotropic phase (the ISO2) at high $c_{\mathrm{DNA}}$. The ISO2p and ISO2 may or may not be the same phase.

However, by far the most striking and unexpected feature of the GCCG phase behavior is the large hatched ($c_{\mathrm{DNA}}$,$T$) area in ***Fig. 1B***, a broad range of temperatures at intermediate concentrations, in which a DNA phase is found that is at the same time crystalline and optically isotropic, a combination that is absent among the known phases of NA oligomers and polymers in solution. DNA crystals, such as those of the Drew-Dickerson dodecamer, for example [23], are built on the close packing of parallel extended WC duplex strands, and therefore have uniaxial or lower symmetry and are birefringent. Since this new phase is so unusual for single component DNA, we preview its structure in ***Fig. 2***, as a guide for the detailed discussion of the experiments, evidence, and modeling leading to its discovery. X-ray and optical study of this isotropic GCCG phase, which we term the BCCX, shows that it is a three dimensionally (3D) periodic liquid crystal structure having the spatial symmetry of a body centered cubic crystal lattice. The BCCX lattice consists of octameric clusters of G's at the cube corners and center, connected into a network by WC duplex rods (struts) along the cube diagonals, each strut made of six GCCG molecules. The resulting cubic lattice is an open network structure, with a DNA volume fraction of ~0.2 (BCCX phase concentration $c_{\mathrm{BCCX}}$ ~ 395 mg/mL, including the $Na^+$ native counterions). This makes the BCCX the least dense of any of the LC phases it can coexist with in solution, enabling it to be readily separated by centrifugation.

This phase first appears upon extended incubation, over periods of ~1 day and longer, of GCCG samples in the hatched area of ***Fig. 1B***. As $c_{\mathrm{DNA}}$ is raised above ~450 mg(ml) with NaCl regulated to around 100 mg/mL, exhibited the growth of a population of isolated domains into existing NEM/COL textures. These scattering domains initially show up in the x-ray scattering as a distribution of localized resolution-limited bright spots in SAXS and WAXS images (***Fig. 3A***), indicative of a randomly oriented distribution of what appear to be small single crystallites. The number and size of these crystallites increases with incubation time, and, as this happens it becomes clear that these spots are localized onto a set of rings, such that, as the volume fills with crystallites, the scattering evolves to give an x-ray powder pattern of rings (***Fig. 3B***). This pattern of scattered intensity is circularly averaged around $\boldsymbol{q} = 0$ to obtain the powder diffraction data in ***Figs. 3-6***. The hatched area of ***Fig. 1B*** shows the range of stability of this powder scattering pattern once established, indicating, for example, that the underlying structure has high thermal stability. ***Fig. 4*** shows that this powder pattern exhibits a set of reflections that can be accurately Miller-indexed to the scattering expected from a body-centered cubic lattice of lattice parameter $a = 65.99$Å (space group #211, I432). The table shows that the calculated reciprocal lattice wave vector magnitudes $g_{(j,k,l)}$ ($\boldsymbol{g}_{100} = 0.0952$ Å$^{-1}$) agree very well with their corresponding experimental values $q_{(j,k,l)}$, for electron density wavelengths $\lambda$ down to $\lambda \sim 10$ Å, about 15% of the unit cell dimension.

As the BCC crystals are first forming at low $c_{DNA}$ ~400 mg/mL, the BCC appears at multiple overlapping spacings that have nearly the same $q$-values, including diffuse rings with a shorter coherence length, observable in ***Fig. 6***. However, with time and increasing $c_{\mathbf{DNA}}$ the final crystal phase was found to have this BCCX structure of ***Fig. 2*** over the entire hatched area, remarkably with a lattice parameter that exhibits no significant dependence on either $c_{DNA}$ or $T$ in this region, as illustrated in ***Figs. 5,6***. The BCCX is resilient to temperature, retaining its structure to $T \sim 80$°C where it melts directly to the structurally disordered, and also optically isotropic, ISO phase. The ISO phase scattering $I(q)$ is featureless at $T = 92.2$ ºC (***Fig. 5***), but scattering immediately returns upon cooling to the NEM phase, or at higher concentrations the COL phase, once again exhibiting the diffuse peak at $q \sim 0.2$ Å$^{-1}$ ($T = 25.4$ ºC in ***Fig. 5***). ***Fig. 5*** reveals the hysteretic behavior of the BCCX, with the crystal failing to reappear through the process of supercooling the ISO by 60ºC. Even more extreme hysteresis is found, in that at room temperature it can take many days for the BCCX peaks to reappear. ***Fig. 6*** shows that near room temperature the lattice disappears in a phase transition for $c_{DNA} \gtrsim 700$ mg/mL, the x-ray scattering becoming dominated by a diffuse ring at $q \sim 0.2$ Å$^{-1}$, indicative of a phase transition to the positionally disordered ISO2 phase, which DTOM experiments show to also be optically isotropic. This combination of isotropy and short range order appears to be similar to that found in the high-$c_{DNA}$ isotropic phase of GCCGp. [13].

Isolated growing crystallites pass through the stage shown in ***Fig. 3*** where they exhibit diffraction limited single crystal Bragg scattering that in many cases illuminates only single pixels in the scattering plane, requiring coherent scattering and defect-free BCC ordering of crystals of thousands of unit cells . Discernable Bragg peaks persist in the powder averages only out to $q$ ~1.2Å$^{-1}$ (electron wavelength $\lambda$ ~5Å), indicating that few-Å scale molecular position fluctuations must be reducing the peaks at larger $q$. Even the base stacking peak at $q$ ~2Å is broadened, typically being considerably narrower and more prominent even in columnar LC DNA. Nevertheless the BCCX is a crystal phase, a relative of the blue phases - arrays of topological defects in a chiral nematic liquid crystal director field that is everywhere fluid, spontaneously appearing and organizing into a cubic lattice. They are crystals without any atomic-scale positional order.

The lack of concentration dependence of the BCCX in ***Fig. 5*** can be usefully contrasted with that of the NEM and COL phases of GCCG presented in ***Fig. S11*** and also visible in ***Fig. 6***, which shows that as concentration $c_{DNA}$ increases, $q_p$, the position of diffuse scattering near $q_p \sim 0.2\,Å^{-1}$, indicating that the sample contains a residual fraction of coexisting COL phase, also increases (blue crosses). This phenomenon reflects the fact that upon increasing $c_{DNA}$, the double helical columns, incompressible in length, get closer sideways. This behavior is found in long duplex DNA [24,25], complementary dodecamers [7] and shorter oligomers [26].

Reduction of intercolumnar distance can be produced by controlling either the DNA concentration or the DNA osmotic pressure directly [**Error! Bookmark not defined.**,27], the latter showing that *h* data can used to read out DNA osmotic pressure. Accordingly, by showing the growth of density in the coexisting COL phase, data in ***Fig. S11*** demonstrate that BCCX structure withstands the increasing osmotic pressure, a behavior coherent with its crystalline structure. From the intercolumnar distance, by a simple construction (SI S***ection SI3***) we estimate that the osmotic pressure, $P$, in the investigated interval to range from $P$ ~2 to ~5 MPa over the range 400 mg/mL $\lesssim c_{\mathbf{DNA}} \lesssim$ 700 mg/mL. Over this range ***Fig. 6B*** shows a fractional compression of the BCCX lattice of $s = \delta a/a < \sim 0.3\%$, and therefore an estimate of the BCCX lattice bulk modulus is $B \sim P/s \sim 10^9$ Pa = $\sim 10^4$ atm.

***Model***

We considered the modes of NA self-assembly which could lead to a low density, nearly incompressible, crystalline structure, starting from the following observations: (***i***) Since the lower temperature region of the BCCX phase in ***Fig. 2*** is occupied by the columnar phase, we assume that the basic self-assembly motif of GCCG will be into G-C complementary base-paired "brick-work" duplexes. Such fiber-like columnar aggregates create the mechanism by which rigid, low density structures can be made; (***ii***) Such duplexes of finite length terminate with GC tails at the two ends; (***iii***) The ability of G to interact with monovalent cations to form G4 structures is very well known [28]. Stacked, two or three-layered G4 offer the possibility of linking 8 or 12 duplex assemblies by coupling their tails together. It is important to note that the native $Na^+$ cations are present in sufficient quantity in GCCG solutions ($c_{Na+}$ = 1.34M/L at the BCCX concentration) to render the DNA non-acidic and, with the three phosphate groups in GCCG, there are always enough cations around to form every G in the material into a quartet; (***iv***) Side-by-side intercolumn interactions must be largely irrelevant in the determination of crystal structure and lattice parameter. (***v***) The lack of dependence of BCC lattice parameter, and therefore the BCCX structure on solution concentration $c_{DNA}$ or temperature is indictive of a nearly fixed BCCX concentration, $c_{BCCX}$. (***vi***) Centrifugation experiments indicate that the density of the BCCX phase, $c_{BCCX}$ is less than ~450 mg/mL, which is much less than the solution concentration $c_{DNA}$ over most of its concentration range.

The simplest models for the BCCX incorporating considerations (***i***) to (***vi***) involve the formation of 3D open frameworks built by connecting finite-length WC base-paired duplex rods (struts) by nodes formed from groups of CG tails at their terminal ends through G4 motifs as in ***Figs. 2,7***. We have constructed what we consider to be an exhaustive set of such models, subject to the constraints imposed by the XRD-indicated BCC crystal structure, the optical isotropy, and the measured BCCX lattice parameter *a*. In these models the nodes are centered at points of cubic point group symmetry, having on average the point symmetry elements of an octahedron. The instantaneous electron density distribution within a G4 node at any given time will be of lower symmetry than this, but the average requirement for node structure can be most readily modeled by making them spherical and of uniform density, as the simplest approximation. The resulting lattices are shown in ***Fig. 7***, wherein each strut comprises a WC paired region of length *s* that inserts its CG terminations into nodes of radius $r$ at either end, the total length of the strut being $d = 2r + s$, where *d* is the node center separation. In the case of struts parallel to the cube edges or parallel to the diagonals we have $d = a$ or $d = (\sqrt{3}/2)a$, respectively. *r* and *s* depend on the valence N of each node, i.e. the number of GCCG terminals converging in it, and on the number M of GCCG oligomers present in the strut, respectively. *r* can be estimated assuming the node to be spherical and completely filled with DNA of uniform density $\rho_{DNA}$

$$r = \left(\frac{3Nm_{GC}}{4\pi\rho_{DNA}}\right)^{\frac{1}{3}}, \qquad (1)$$

where $N$ is the number of GC units in a node, $m_{GC}$ is the mass of a single-stranded GC element, (about ½ the mass of a full GCCG oligomer - MW = 1171.8 g/mol) and $\rho_{DNA}$ = 1687 mg/mL, the density of neat GCCG DNA. *s* can similarly be calculated as $s = 3.39(2M - 2)$Å from the known WC inter-base stacking distance that we take that to be 3.39 Å based on our x-ray observations of nanoDNA in B-form WC columnar stacks [**Error! Bookmark not defined.**].

The BCCX can be built by constructing a simple cubic lattice of unit cells of edge *a* and dressing them with a 2-node sites form factor at a cube corner and the cube center, with their connecting strut, in each unit cell. There are three lattices compatible with this node-strut structures and with the BCC symmetry. As sketched in ***Fig. 7***, they feature a node at one corner of the unit cell (UC) and another at the UC center. We jointly varied *N* and *M* to obtain BCC unit cells with $a$ ~ 66Å, and calculated the resulting concentration of the phase in mg/mL, including $Na^+$ counterions. Lattice #1 (L1) places two sets of struts parallel to the UC edges, one running through the UC corner and the other through the UC center, creating an interpenetrated pair of simple cubic lattices, where the cube corners of one lattice are at the cube centers of the second lattice. In L1 the node valence is N=6 and the condition $a$ ~ 66Å yields M=8. L2 places the struts along the diagonal lines between the corner of the UC and the center of the UC. In L2, N=8, a structure compatible with a two-layered G4 assembly and M=6. L3 merges these two models, with struts along the edges of the UC as in L1, but also along the diagonals of the UC, as in cell L2. In L3 the node valence is N=14 and the strut length is either M=6 or M=8 depending on their direction with respect to the BCC axes. ***Fig. 7A-D*** and the table summarize the characteristics of these lattices. L1 and L2 give concentrations $c_{BCCX}$ = 351 and 396 mg/mL, respectively, both satisfying the condition of consideration (***v***) above, whereas L3, due to its much higher concentration, can be eliminated on this basis. In L1 and L2, with 48 GCCG oligomers per UC volume of $a^3$, the native $Na^+$ ion concentration in the aqueous solvent of the BCCX crystal is 1.34 M/L. Condition (*iii*) enables discriminating between L1 and L2 since only the latter has a node valence multiple of 4.

We further examined the nature of such BCCX lattice assemblies using atomistic molecular dynamics simulation. Run times available for atomistic simulation of the BCCX system were not of sufficient duration to model large-scale collective behavior, for example crystal self-assembly starting from a completely disordered state, but could be effectively used to probe structure and fluctuations within a node under circumstances typical of its tethering by the struts in a BCCX lattice. To this end, we carried out simulation runs starting from the state shown in ***Figs. 7,S4***, in which a BCCX network element is built using 48 5′-GCCG-3′ oligonucleotides. These are arranged to form eight B-type DNA duplex struts each six oligonucleotides in length, all terminated at one end on a single common node. We searched in the literature for nonplanar H-bonding modes of the Gs that might enable a node assembly with octahedral symmetry but did not find suitable candidates, so we settled for eight G′s in a standard pair of stacked quadruplexes with two sodium ions as a starting condition. In the BCCX both 3′ and 5′ ends populate the nodes, but for this simulation all of the struts have their 5′ terminations in the node. The eight struts were arranged to radiate out toward the corners of a cube centered on the G-quartet pair, forming an eight-armed star with the node at the center, as shown in ***Figs. 8A-C***. This starting assembly was immersed in water with the native sodium counterions, shown explicitly in ***Fig. S9***, then simulated in X, Y, Z periodic boundaries using the Amber force field on the Gromacs simulation package. This gave a simulation containing 48 GCCG oligomers, 13990 water molecules and 144 sodium cations (48119 centers total) with initial dimensions of the periodic box of 8.4 x 8.4 x 6.9 nm. During the 150 nsec simulation the configurations of this system were restrained by 8 harmonic springs of constant ~5.5 kJ/mol*$nm^2$, each pinning an O6 oxygen atom on the guanine terminating the free end of a strut, effectively keeping those 8 atoms near the corners of a cube equal in size to the BCCX unit cell obtained from the XRD, emulating their behavior in the BCCX lattice. Simulations were carried out according to the procedure in the *Supporting Information*.

We note a number of structural features of this system seen during 150 ns of simulation time: (***i***) WC paired portions of struts remain mostly intact with nearly B-form helicity as seen in ***Fig. S10***; (***ii***) Soon after the start, typically for t $\lesssim$ 10 ns, one of the two initial Na+ ions routinely escapes from the core grouping, which can be seen by comparing ***Fig. 8C*** with ***Figs. 8E,F***, most likely a result of the electrostatic repulsion between those Na+ ions in a stack of only two; (***iii***) The two initial G-quartets in the quadruplex, as is *Fig. 8C*, undergo a fission event, each splitting in half where half of each recombines into a single persistent G-quartet around the single remaining sodium, which appears to form the foundation of a long-term hydrophobic core ***Fig. 8E,F***; (***iv***) The G-quartet present at $t = 150$ nsec has two G-bases from each of the initial G-quartets, with the trapped Na+ atom seen to wander up and down the axis of the hydrophobic core, facilitating swapping of new, unliganded G-bases into and out of the active G-quartet; (***v***) G-bases that are not liganded into the active G-quartet remain plastered to the surfaces of the G-quartet, forming an extended, non-specific hydrophobic cluster as seen in ***Figs. 8E,F*** and videos in Supplementary ***Section S4***; (***vi***) The final active G-quartet fluctuated around a specific final structure in which its plane is oriented into the common base-pair plane of a pair of struts along a cube diagonal and on either side of the plane of that G-quartet, forming a continuous hydrophobic column that tapers from the quartet into the struts, as seen in ***Fig. 8D*** and in Supplementary ***Section S4***; (***vii***) Additional struts that abut the node obliquely generally extend their single-stranded regions into the persistent hydrophobic core associated with the G-quartet and otherwise bury the end of their WC paired stack into either the side of the core or into a neighboring strut’s major or minor grooves as seen in ***Fig. 8D*** and the videos of Supplementary *Section S4*; (***viii***) C-bases from the single-stranded overhangs of the struts also are seen to stack into the hydrophobic core around the foundational G-quartet, as seen in ***Figs. 8E,F;*** (***ix***) At least one G-base from the recessed 3’-terminal of a WC duplex strut region has been seen to undo its WC base-pairing into order to participate, unpaired and unliganded, in the stacking around the hydrophobic core nucleated on the persistent G-quartet, as reflected by deviation of strut 7 from typical helicity in ***Fig. S10***; (***x***) Ambient Na+ cations visit the entire volume of the simulation box, but appear to be more frequently collected around the DNA structures of the GCCG L2 node (as seen in *Fig. S9*); (***xi***) There is no swapping of cations into or out of the node core since the ion present there at the end of the simulation is the same as one of the two that started at that location; no other cations form as specific or persistent an association with the DNA structure.c

*Discussion*

A rough estimate of the relative stability of the COL and BCCX phases can be made by considering the free energy change $\Delta G$ produced upon introducing breaks in GCCG duplex chains, generating free ends with unstacked and unpaired Gs. In the case of L2 this involves making four such breaks and then inserting the 8 terminal Gs into node balls, within each of which they form a two-layered (2L) G4 stack. The free energy gain $\Delta G_{2LG4}$ involved in the formation of the 2LG4, must overcome the loss for the 4 breaks, each of which involves the loss of 3 nearest-neighbor stacking contribution $\Delta G_{NN}$, that we can evaluate using the conventional tools for the thermal stability of DNA duplexes [29]. In the evaluation we should also consider the 2LG4 stabilizing contribution $\Delta G_{DE}$ due to the "dangling end" stacking of the unpaired C to the WC-bond portion of the struts:

$$|\Delta G_{2LG4}| > |12\Delta G_{NN} - 4\Delta G_{DE}| \approx 4 \times \left(3 \times 2.3\,\frac{kcal}{mol} - 0.75\,\frac{kcal}{mol}\right) - 8 \times 0.5\,\frac{kcal}{mol} \sim 20\,\frac{kcal}{mol}, \qquad \text{Eq. 1}$$

where the second term in parenthesis is the correction for salt concentration [29]. The resulting $\Delta G_{2LG4} > 20$ kcal/mol is compatible with the free energy measured for the assembly of two layered G4 [30,31], although it should be noted that the 2LG4 structure we are here considering differs from the ones used to determine the free energy available in literature, based on G4 forming within oligomers containing repeats of guanosines. We suggest that the free energy associated to the BCCX 2LG4 formation could be larger than the literature values, since the association of 8 chemically independent chain terminals, unconstrained by chain loops as in the G4-based secondary structures of oligomers, benefits from a larger conformational freedom, enabling guanosines to relax to their position of strongest stacking. Indeed, $\Delta G_{2LG4}$ should not only satisfy the necessary stability with respect to the COL phase, expressed in Eq.1, but also provide the free energy required by the conformational tightening associated to the BCCX lattice, in which fluctuations are certainly quenched with respect to the COL phase with equal $c_{DNA}$. This notion combines with the observation that the BCCX structure is not found in GCCGp, where the terminal phosphates destabilize the 2LG4 nodes by adding an element of electrostatic repulsion, and with the fact that BCCX does not appear in conditions of lower or larger ionic strength, both suggesting that the stability of BCCX is marginal, where the free energy gained in forming the 2LG4 nodes is entirely spent in the stabilization of the BCCX structure.

We argue that conformational tightening is also at the origin of the large thermal stability of BCCX, whose melting is 30 degrees above that of the COL phase despite the smaller $c_{DNA}$. The stability of BCCX mainly depends on that of the WC struts, being these the weakest bonds in the structure. Thus, the huge increment in their stability with respect to the COL phase must be a consequence of a increment in the free energy associated to the GCCG WC bonding that can be attributed to a reduction of the entropy penalty for their formation, in turn suggesting a significant reduction of bond-breaking fluctuations of the BCCX struts with respect to the COL columns.

The L2 node within the MD simulation appears to have a very simple, dynamic structure where an underlying G-quartet built on a single trapped sodium cation forms a nucleus upon which other nearby hydrophobic moieties can collect. The presence of this cation quite clearly confers an organizing effect on colocalized G-residues within the simulation and its migration can shuffle which Gs are organized as such. The structure is stable to at least $t = 150$ nsec and is possibly not the end state of the assembly; coarsening dynamics in observed GCCG LC samples are on a time scale of days to weeks, which is completely inaccessible in an MD simulation confined to microseconds at the longest. The overall disorder of the node *in silico* suggests that it can escape the requirement for octahedral symmetry by simply being highly flexible and disorganized, thus averaging out its anisotropy during XRD. The apparent lack of overall preferred multiplicity in the simulation suggests that such a simple structure could easily be formed within conventional GCCG nematic and columnar phases, creating the LC texture attenuations noted in ***Figs. S12,S13*** in addition to being able to form the BCCX in its ultimate expression. It is also possible that the extended conformation of the single-stranded regions around the core at the start of our simulations biased against trapping two cations within; if hydrophobic dams are formed on G-quartet surfaces in lower order NEM or COL systems, stable quartets tethering 4 struts apiece may be able to encounter one another in a way where the quartets can assemble with each other into nodes with a larger number of stably trapped ions.

The observation of the BCCX phase brings with itself the two questions of why it is present in the GCCG phase diagram only, never having been observed in our previous studies of several hundred LC-forming DNA oligomers [7-17], and why the periodicity adopted in BCCX involves 6 GCCG molecules in the WC bonded struts. While we do not have a definite answer to the question, we argue that a key element to be considered might be that given the GCCG duplex helical periodicity, the WC bonded segment of the BCCX strut is close in length to a complete turn of the B-DNA helix. This enables the possibility that each strut-end would approach coupling into its G-aggregate with the same average orientation relative to the crystal lattice directions. We argue that any correlation among the azimuthal angles of the WC duplexes at the 2LG4 nodes would tend to favor struts of length equal to a multiple of the WC B-DNA helical pitch. Exploration of the orientation correlation of the WC pairs closest to the node in the MD simulation is perhaps not conclusive; the highly disordered structure seen *in silico* may not actually be the ending state for the BCCX node given that real world coarsening times are so much longer than the simulation. The flexibility of the node structure seen in simulation could easily favor a broad range of interaction modes and did assemble with the surrounding struts in a way that blended with duplex continuity at later time points [*Fig. 8D*]. At the same time, longer WC struts would have markedly reduced thermal stability since in larger BCCX unit cells the crystal-induced structural stability and fluctuation reduction would be much less. Thus M=6 could be the sweet spot combining short WC aggregate with a complete helical turn.

Along these lines, a good candidate for exploration would seem to be "sDD", the12-base oligomer of sequence GCGCGCAATTGC [32,33] that is self complementary with GC overhangs, and whose hybridization leads to chemically continuous structure of length and terminals matching those of the M=6 GCGC struts. However, our limited studies to date of sDD have not revealed a BCCX phase. This could very well be a consequence of the kinetic barrier involved in the formation of such a phase. Indeed, even in the weakly bonded, highly dynamic GCCG solutions, nucleation of BCCX encounters a significant barrier. Given that the local melting attempt rate should be much higher in GCCG than in sDD, sDD-BCCX nucleation might be many orders of magnitude slower.

***Conclusion***

While there is an ample literature reporting the self-assembly and secondary structuring of NA oligomers based on WC or G4 binding modes, examples of structures based on their combination are missing. The nanoporous crystalline BCCX scaffold adopted by GCCG as its favored equilibrium state in a large range of its phase diagram is a rare example of a mode of association crucially depending on both. We find BCCX to form via nucleation in solutions already collectively organized through WC pairing. In these, G-C base-pairs are built quickly given a rate equation that involves only two components, but they are not as stable, while G-multimer structures, like G-quartets, are highly stable but slow given a rate equation that may involve eight parts at once. Base-pairs formed early can be reversed slowly over time and replaced by G-quartet-like associations to create structures that contain both regions of WC base-paired helix and stacked G-quadruplex. Terminal ends that are involved in quartet or quadruplex become disabled for involvement in G-C pairs.

It is intriguing that the smallest nanoDNA oligomers also tend to have the most complicated phase diagrams in the DNA family. Indeed, longer oligomers tend to have a definite preference for WC vs. G4 mode of assembly. Longer complementary oligomers are more prone to exhibit long-DNA-like columnar behavior. Oligomers with G repeats typically fold around G4 nuclei. The 4mer GCCG is small enough that the WC-G4 mixed-mode assembly behavior can happen in a semi-repeatable manner, unlike what may occur in longer oligomers with high GC content and many more molecular degrees of freedom, where the association might produce a much larger variety of structure that may have a less repeatable or perhaps have a more dead-end effect on the phase diagram of that material.

The combination of multivalent nodes and linear connectors is the basic design for soft, porous, three-dimensional reversible networks in soft matter, often appearing in the form of hydrogels. Such networks are typically formed by the association of complex units that control the valence and the internode-spacing, or they result in very disordered structures [34]. Only by highly complex monomer design these networks can achieve long range periodicity [5]. The GCCG self-assembly reported here expands the realm of soft matter and DNA nanoscience to include periodic, nanoporous networks by self-assembly in systems of single component, and ultra-small, molecules. The WC strut and G4 node motif introduced appears to have potential for use as a general structural theme for the self-assembly of mesh-like framework nanostructures.

*Acknowledgments*

This research was supported in part by the NSF Biomaterials Program under Grants DMR-1611272 and DMR-2005212, by NSF MRSEC Grant DMR-1420736 to the University of Colorado Soft Materials Research Center, and by the NIH/CU Molecular Biophysics Training Program to GPS. This research used beam line 11.0.1.2 of the Advanced Light Source at the Lawrence Berkeley National Laboratory supported by the Director of the Office of Science, Office of Basic Energy Sciences, of the U.S. Department of Energy under Contract No. DE-AC02-05CH11231.

*Figures*

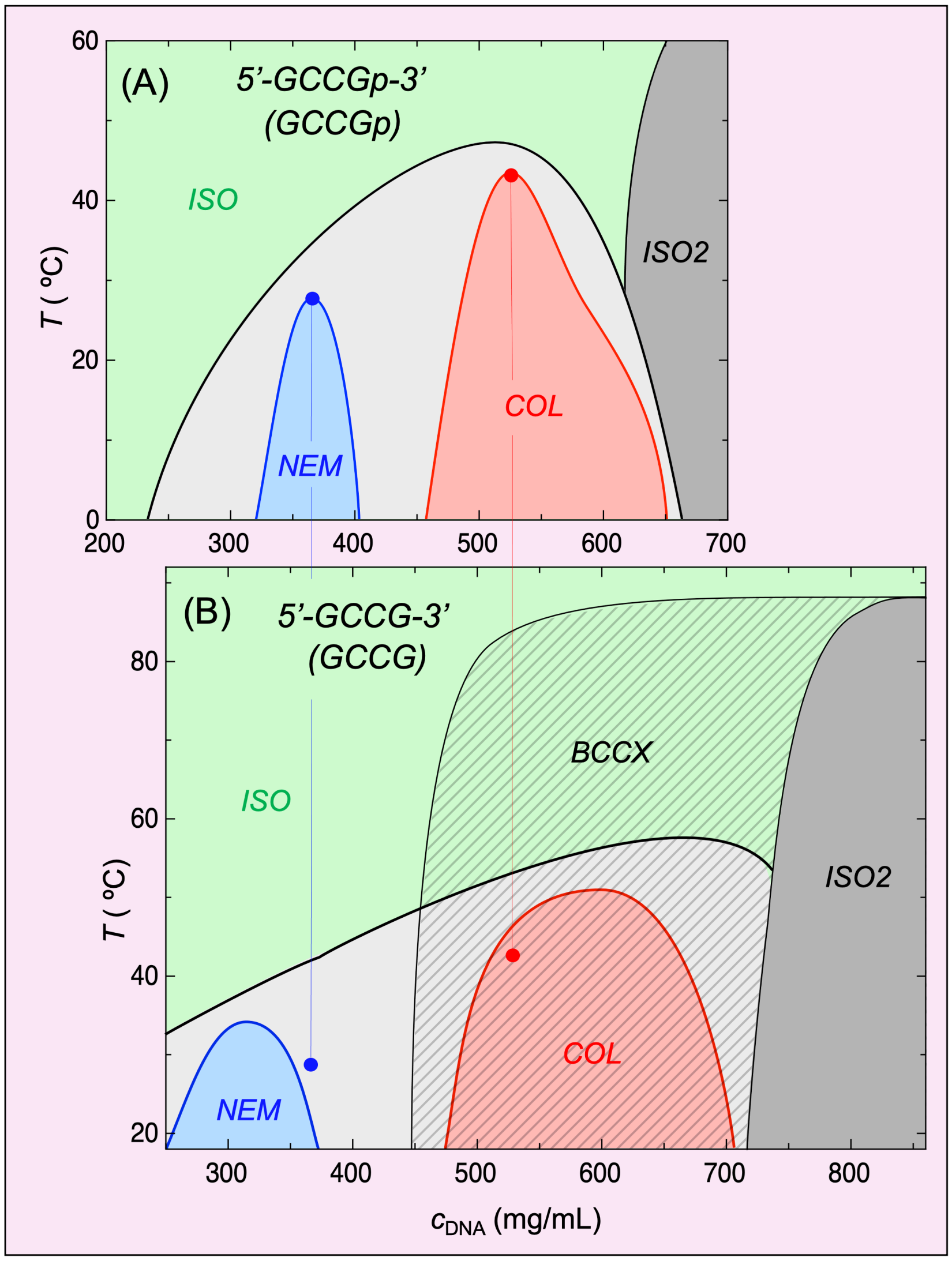

***Figure 1***: Concentration, $c_{\mathrm{DNA}}$, versus temperature, $T$, phase diagrams for G-C DNA 4mers 5′-GCCGp-3′, terminated at 3′ by a phosphate group, reported in [Ref. 12], and 5′-GCCG-3′ terminated at 3′ by an undecorated hydroxyl, reported here. Concentrations range up to 800 mg/mL, which is ~50 weight% DNA. The solid dots indicate, on both phase diagrams for comparison, the highest $T$ of the NEM and COL ranges of GCCGp. (***A***,***B***) Liquid crystal phases observed in both oligomers at these high concentrations are the isotropic (ISO, green), high concentration isotropic (ISO2, gray), nematic (NEM, blue), and uniaxial columnar (COL, red), with regions of phase coexistence among these (white). (***B***) In addition, 5′-GCCG-3′ exhibits a low-density body-centered cubic crystal phase (BCCX), which is broadly coexistent with the other phases in the hatched region (~450 mg/mL < $c_{\mathrm{DNA}}$ < 800 mg/mL $T$ < ~85°C).

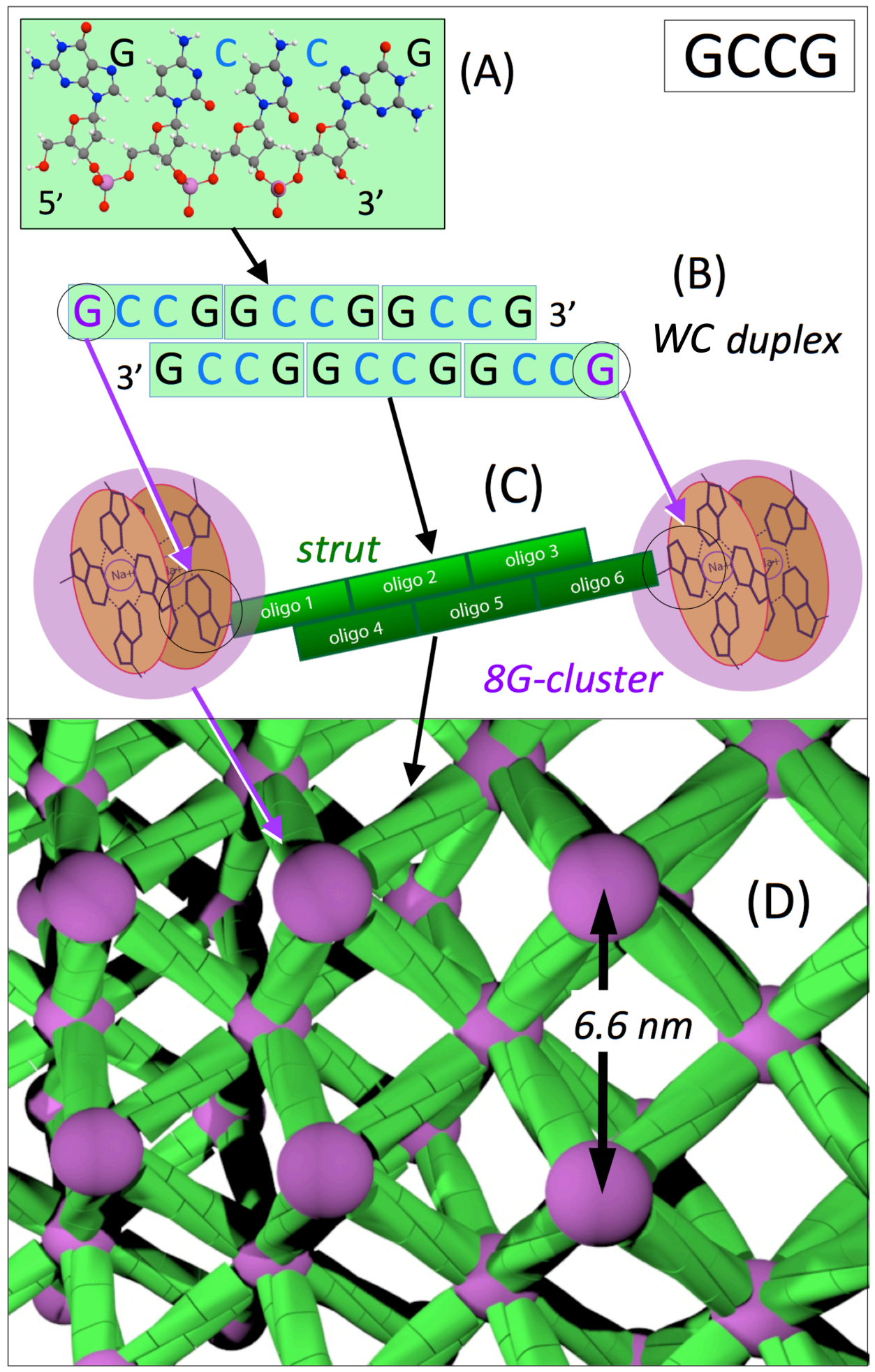

***Figure 2***: View of the hierarchical structural organization of our model of the BCCX phase of 5′-GCCG-3′ (GCCG), introduced as a guide for the detailed discussion of the experiments, evidence, and modeling. (***A***) A schematic 5′-GCCG-3′ oligomer containing Guanine (G), Cytosine (C) and three deprotonated phosphates in the deoxyribose phosphodiester backbone.
(***B***) Oligomers assemble by Watson-Crick (WC) base pairing and stacking into linear duplex columns comprising a brickwork tiling of oligomers, with each duplex pair producing GC overhangs on both its 5′ and 3′ ends. (***C***) The GC-terminated duplex columns aggregate by guanine-to-guanine association into star-like clusters stabilized by the formation and stacking of G-quartets in G-rich nodes. (***D***) Duplex struts of WC base paired oligomers bridge between the nodes to form a periodic BCCX framework phase with struts and nodes on a body centered cubic crystal lattice having a lattice parameter $a = 6.6$ nm, and a DNA concentration $c_{BCCX} = 396$ mg/mL. This drawing gives a qualitative representation of the BCCX electron density in which sub-5Å structure is smoothed out by fluctuations.

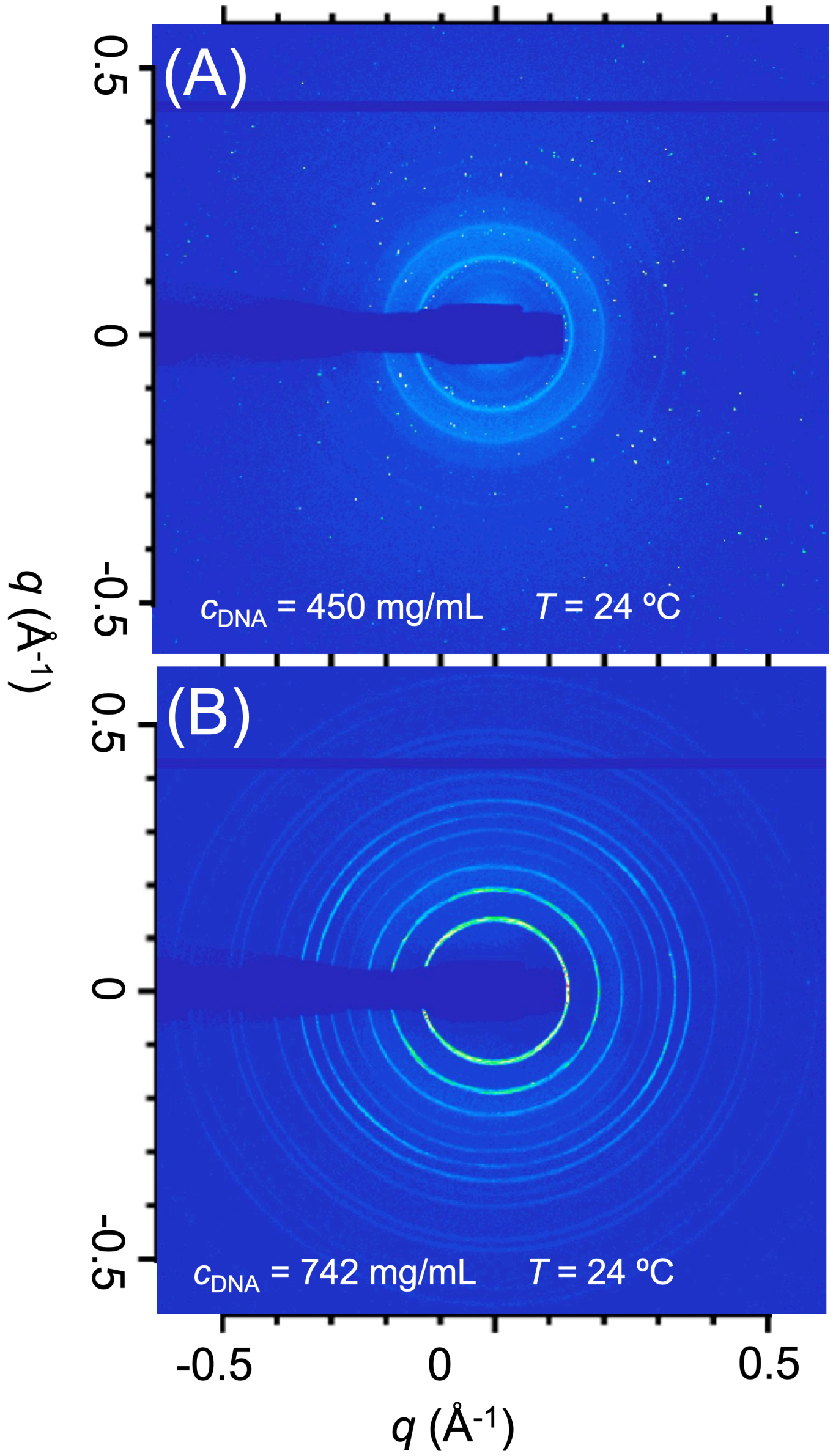

***Figure 3***: The small angle region of 2D WAXS x-ray diffraction patterns obtained from the 5′-GCCG-3′ BCCX phase, forming or having formed from GCCG starting solution exhibiting coexisting NEM and COL phases. (***A***) BCCX growing in as isolated crystallites at $T$ = 24°C and $c_{DNA}$ = 450 mg/mL. At this growth stage scattering from a single crystallite typically illuminates only a single pixel in the scattering plane. This diffraction-limited scattering requires perfect single crystal domains containing a minimum of ~$10^3$ unit cells of 6.6nm dimension – evidence for crystalline ordering. The diffuse rings are produced by both the NEM and COL ordering in the starting solution, as well as BCCX-like structures having only short-ranged ordering. (***B***) Annealed powder scattering pattern obtained at higher concentration. Absence of diffuse scattering indicates that the BCCX phase has filled the sample volume.

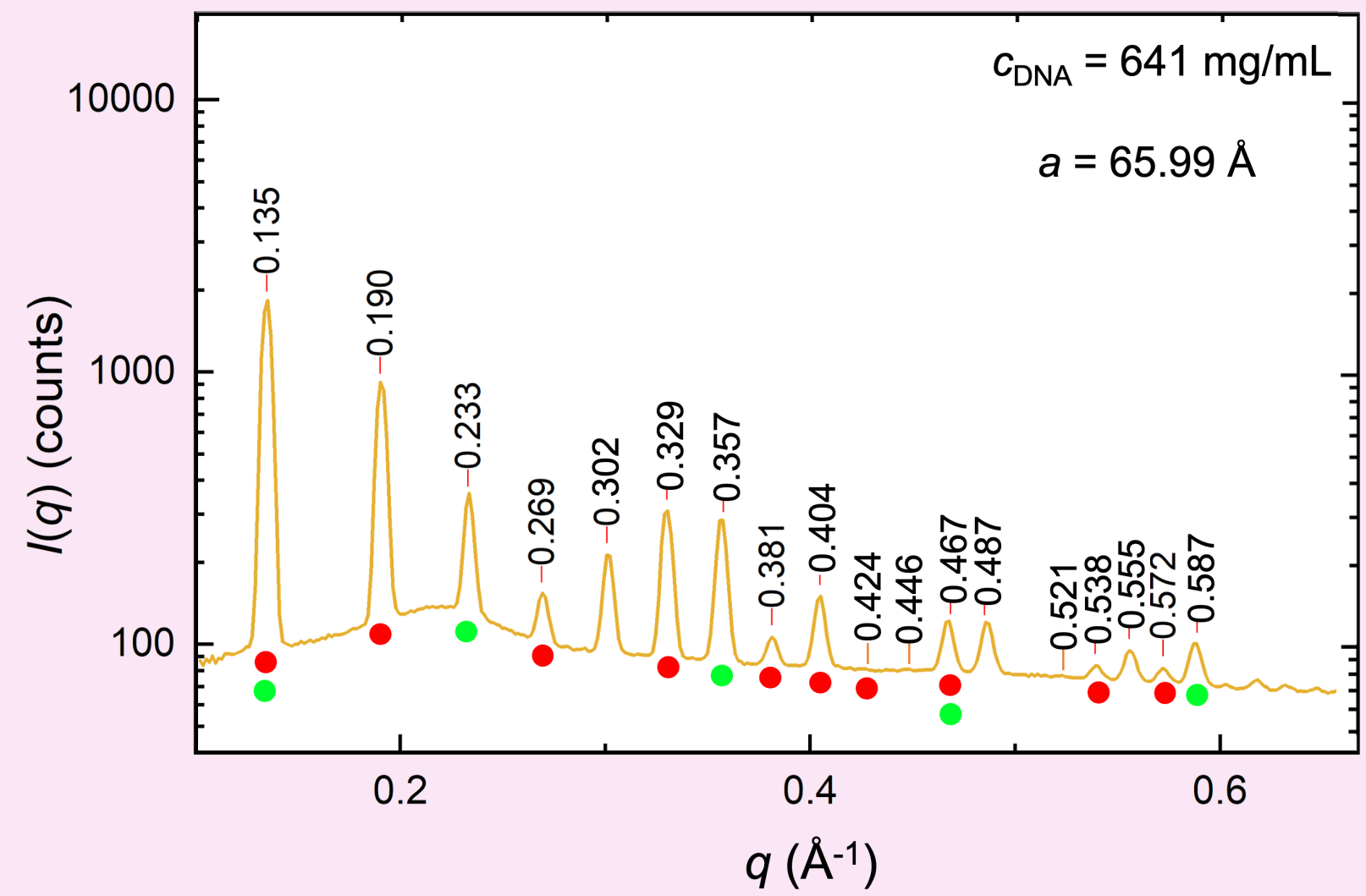

| $q$ (Å$^{-1}$) | index | $(q_{BCCX}/q_a)^2$ | $q_{BCCX}$ (Å$^{-1}$) |
|---|---|---|---|
| 0.135 | (1,1,0) | 2 | 0.13465 |
| 0.190 | (2,0,0) | 4 | 0.19042 |
| 0.233 | (2,1,1) | 6 | 0.23322 |
| 0.269 | (2,2,0) | 8 | 0.26929 |
| 0.302 | (3,1,0) | 10 | 0.30108 |
| 0.329 | (2,2,2) | 12 | 0.32982 |
| 0.357 | (3,2,1) | 14 | 0.35624 |
| 0.381 | (4,0,0) | 16 | 0.38084 |
| 0.404 | (3,3,0) (4,1,1) | 18 | 0.40394 |
| 0.424 | (4,2,0) | 20 | 0.42579 |
| 0.446 | (3,3,2) | 22 | 0.44657 |
| 0.467 | (4,2,2) | 24 | 0.46643 |
| 0.487 | (4,3,1) | 26 | 0.48548 |
| 0.521 | (5,2,1) | 30 | 0.52149 |
| 0.538 | (4,4,0) | 32 | 0.53859 |
| 0.555 | (5,3,0) | 34 | 0.55516 |
| 0.572 | (6,0,0) | 36 | 0.57126 |
| 0.587 | (5,3,2)(6,1,1) | 38 | 0.58691 |

***Figure 4***: Radial scattered intensity scan $I(q)$ obtained by circular averaging a typical powder scattering pattern of the kind shown in ***Fig. 3B***, and its indexing to a face centered cubic (FCC) reciprocal space lattice (body centered cubic (BCC) crystal real space lattice). Diffraction peak locations are annotated in $q$(Å$^{-1}$), and each line in the accompanying table shows a measured $q$-position value, its Miller index assignment on the FCC reciprocal lattice, the ratio of that $q_{hkl}$ magnitude to the simple cubic lattice vector $q_{100} = 2\pi/a$, and the resulting $q_{hkl}$ values calculated for this indexing and a unit cell dimension of $a$ = 65.99Å. The plot and table show a complete 1:1 match of FCC reciprocal lattice points with data peaks in $I(q)$, out to $q$ = 0.6 Å$^{-1}$. Colored spots indicate which peaks would be present for BCC lattices made up from arrays of featureless, infinitely long rods: *RED* – Rods along all of the cubic unit cell edges; *GREEN* - Rods along all of the cubic unit cell diagonals.

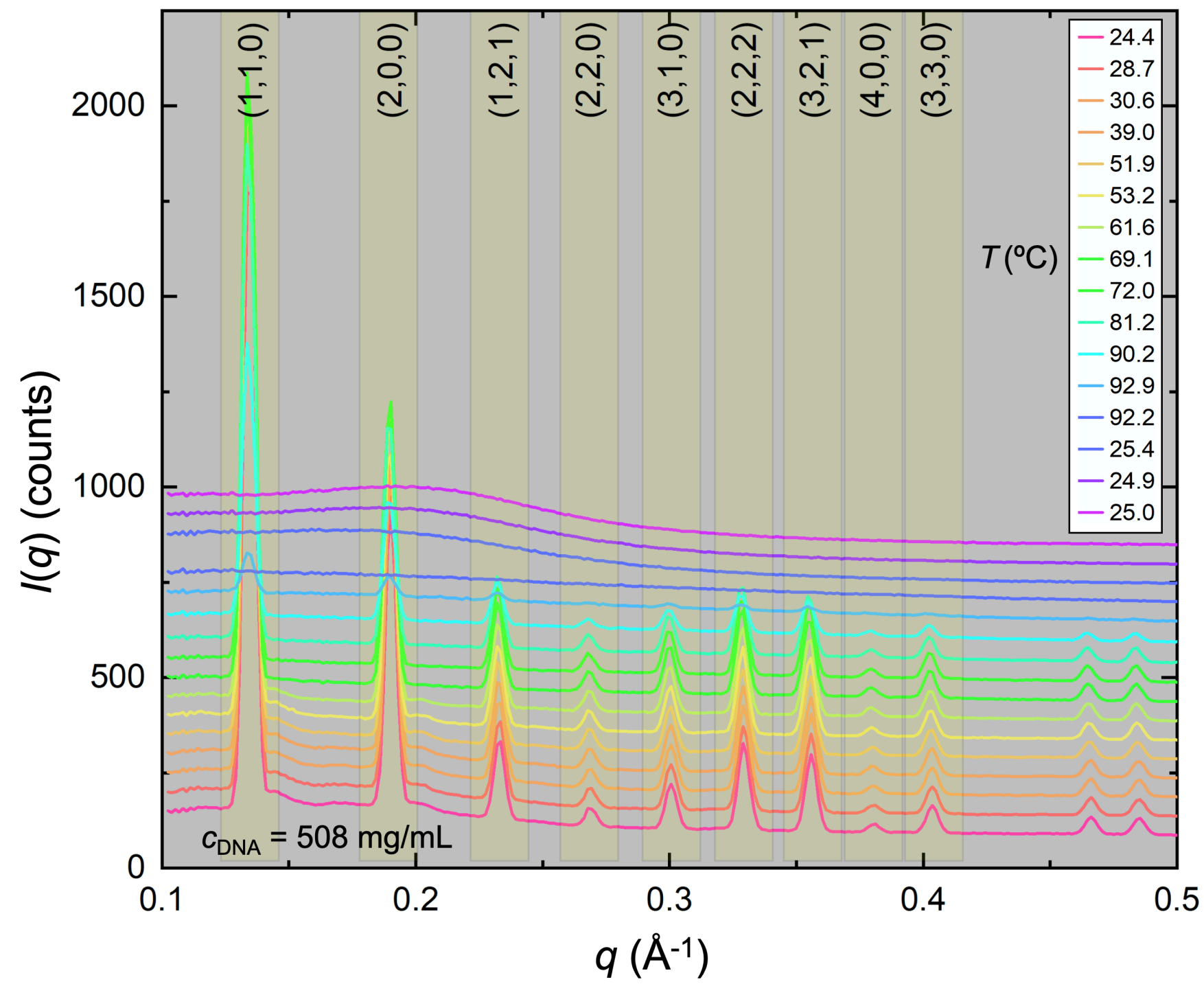

***Figure 5***: X-ray diffraction of the BCCX phase of 5′-GCCG-3′ at $c_{\text{DNA}}$ = 508 mg/mL, for steadily increasing temperature. The Miller indices from ***Fig. 4*** are indicated. The BCCX is seen to persist to $T$ ~ 90 ºC (cyan), melting temperatures that are much higher than those of the duplex columnar NEM or COL phases in GCCG in ***Fig. 1***, or even in other longer DNA oligomers [6-16]. Since the BCCX crystal structure requires intact duplex struts, these high melting temperatures show that the connectivity of the BCCX network structure stabilizes the WC strut assemblies. Upon melting the BCCX x-ray pattern gives way to the featureless scattering of the ISO (blue), but the diffuse columnar peak from side-by-side columnar packing in the NEM phase appears upon cooling, and persists to room temperature (magenta). It may take hours, days, or weeks for the BCCX crystals to reappear.

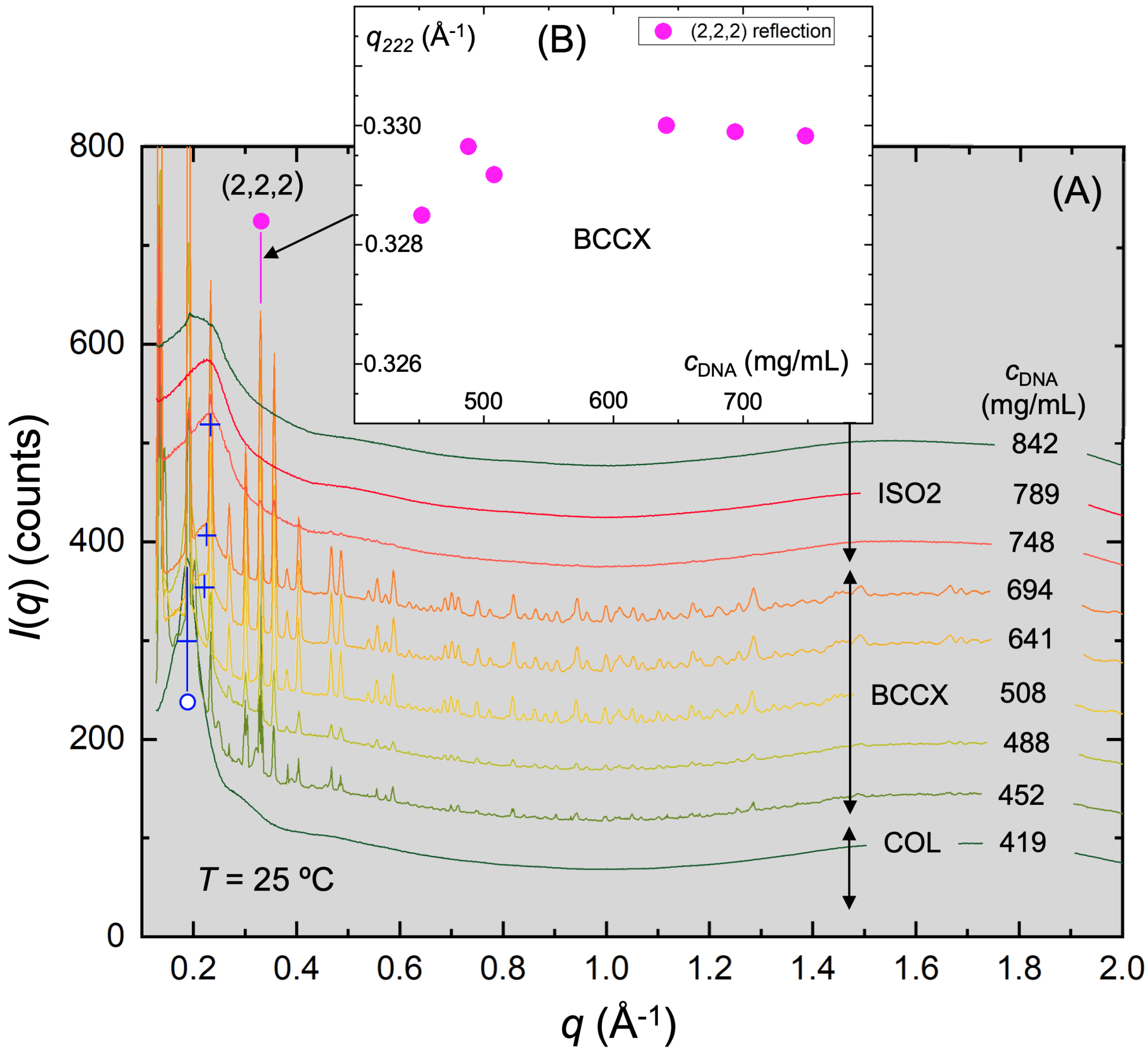

***Figure 6***: X-ray diffraction at room temperature for a broad range of 5′-GCCG-3′ concentrations that includes the BCCX phase. (***A***) Low concentration phases are at NEM/COL coexistence and the BCCX is seen for 450 mg/mL < $c_{DNA}$ < 700 mg/mL, with a coexisting COL phase also in this range (COL peaks denoted by blue +). The BCCX gives way to the ISO2 at the highest concentrations, the latter exhibiting a nematic-like correlation peak at $q$ ~ 0.2 Å$^{-1}$, but without birefringence. In (*A*) there appears to be little change in BCCX Bragg peak positions with increasing $c_{DNA}$. This is quantified in the inset (***B***) which shows fitted positions of the BCCX $q_{2,2,2}$ Bragg peak vs. $c_{DNA}$, yielding a ~0.3% fractional contraction of the BCCX lattice with increasing $c_{DNA}$ over the total BCCX range. The COL peaks (blue+, plotted as open circles in ***Fig. S9***) shift to larger $q$ with increasing $c_{DNA}$, indicating an increase of ~3MPa in the DNA osmotic pressure over the BCCX $c_{DNA}$ range. If it is assumed that it is this compression that produces the 0.3% lattice contraction, then the BCCX crystal bulk modulus can be estimated to be $B$ ~ $10^9$ Pa.

*Figure 7*: Possible node-strut lattice models for BCC lattices that fit the x-ray data. (***A***) Schematic of strut structure, each strut consisting of a WC base paired region with a length s (green) terminated on either end by perfectly spherical, ideally space-filling nodes containing single-stranded segments of Guanine and Cytosine with a radius of $r$ (purple). The node-center to node-center length of a strut is therefore $d = s + 2r$ including penetration into the nodes at either end, such that $s$ depends on both node volume and node-center separation. All lattice parameters are calculated under the premise that the WC regions are duplex DNA stacks with 3.4 Å base pair spacing and that the nodes have an nominal DNA concentration of 1687 mg/mL. The table summarizes the parameters for the specific lattices considered. (***B***) *Lattice model L1* – Nodes have $N = 6$ Gs, struts have $M = 8$ GCCGs and are parallel to the cube edges, forming two inter-penetrating cubic lattices linking cube corners, and the unit cell has 48 GCCGs. (***C***) *Lattice model L2* – Nodes have $N = 8$ Gs, struts have $M = 6$ GCCGs and are all diagonal, linking cube center nodes to corner nodes, and the unit cell has 48 GCCGs. The $M = 6$ struts have WC double helix runs of 10 bases, making the strut length nearly the same as the helix pitch. (***D***) *Lattice model L3* – A combination of lattices L1 and L2 where nodes have $N = 14$ Gs, diagonal struts have $M = 6$ GCCGs, edge struts have $M = 7$ GCCGs., and the unit cell has 96 GCCGs.

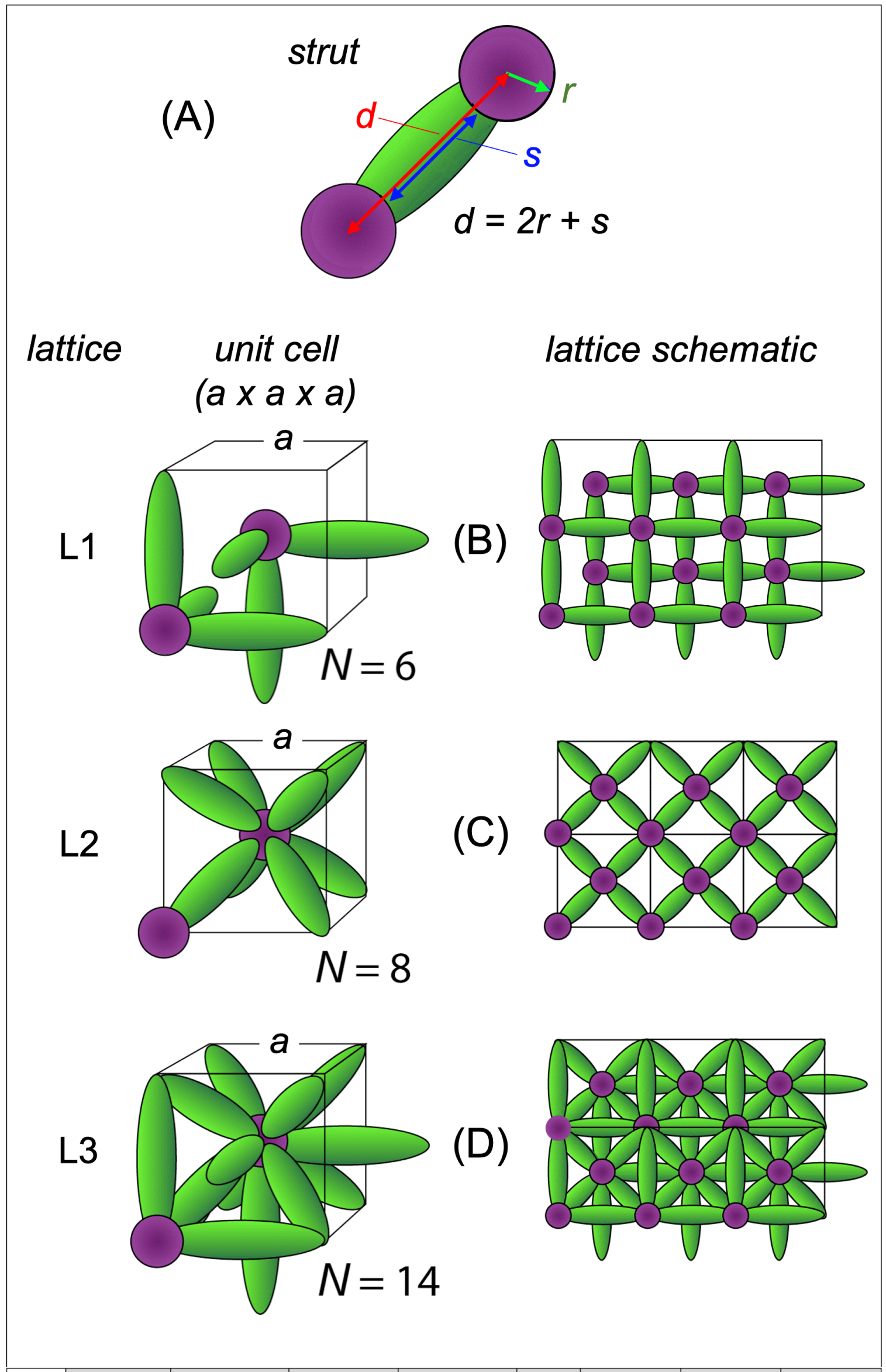

| | $N$ (struts) | $M$ (oligos) | # (oligos) | $c_{DNA}$ (mg/mL) | $a$ (Å) | $q_{110}$ (Å$^{-1}$) | $2r$ (Å) | $2r_{exp}$ (Å) |
|---|---|---|---|---|---|---|---|---|
| L1 | 6 | 8 | 48 | 341 | 66.3 | 0.0949 | 18.8 | 18.5 |
| L2 | 8 | 6 | 48 | 396 | 63.0 | 0.0997 | 20.6 | 23.6 |
| L3 | 14 | 6/7 | 90 | 658 | 65.6 | 0.0958 | 24.9 | 25.2 |

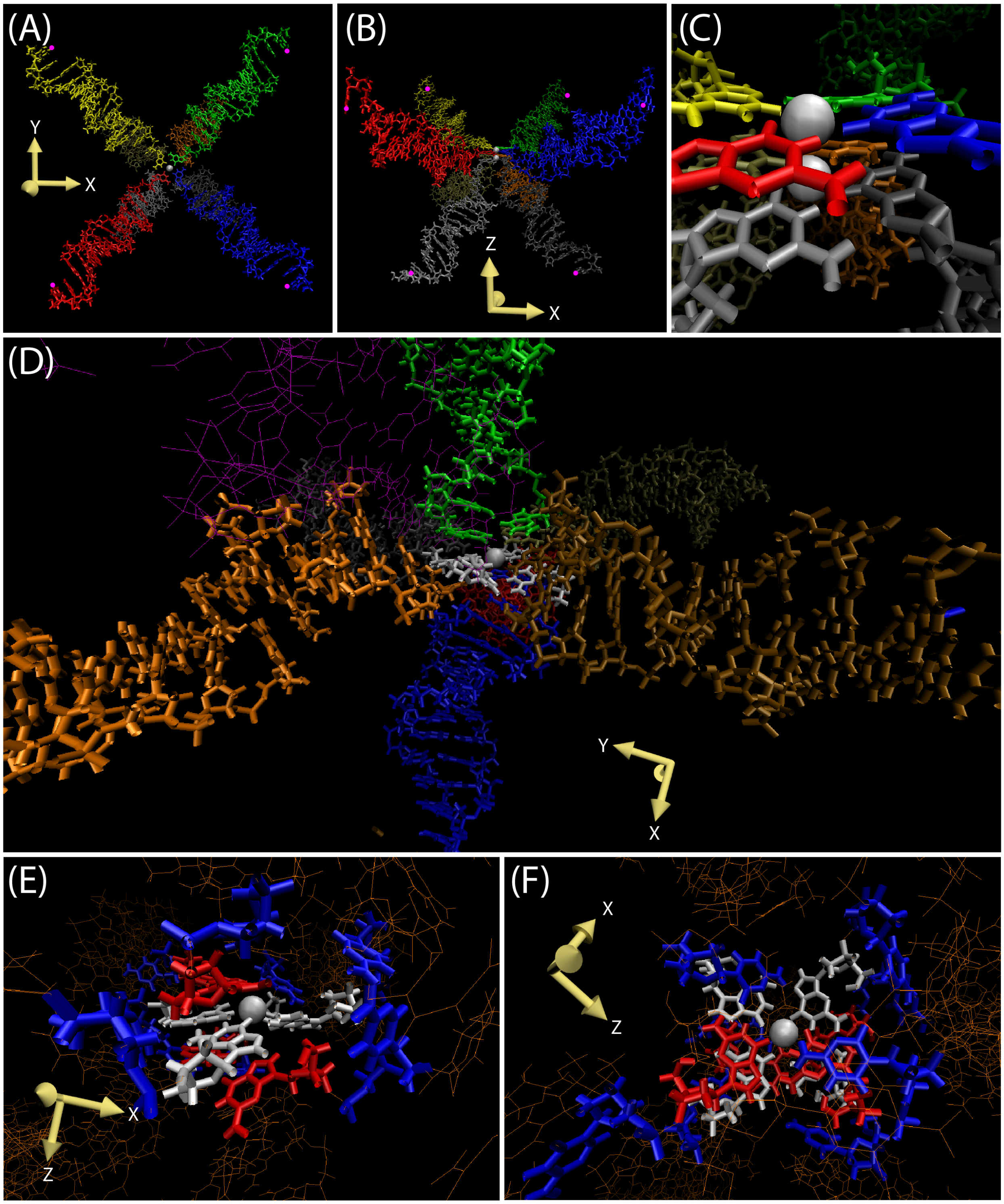

*Figure 8*: Configurations from molecular dynamics (MD) simulation of an eight-arm star representing a single node of an L2 lattice, formed by an $N$ = 8G core, with eight $M$ = 6 struts projecting out toward the simulation cell corners. (***A***) The starting geometry of the simulation is based on a binary stack of two G quadruplexes. This view looks in normal to the plane of the two G-quartets. Oxygen atoms at the strut-termini that are anchored with harmonic restraints during simulation are marked by pink balls. (***B***) The starting geometry looking in parallel to the plane of the G-quartet stack. (***C***) Focused view of the G-quartet stack in the node with the initial two Na+ atoms (white). (***D***) The state of the system at $t$ = 150 ns, with a lone G-quartet and associated Na+ atom (gray) surrounded by WC paired struts (blue, green, orange, brown in front, gray, silver red in back, magenta reduced to wireframe for visibility.) (***E***) A zoomed view of the node at $t$ = 150 ns looking in parallel to the sole G-quartet and Na+ atom (gray) with unpaired G residues (red), single-strand C residues (blue) and the continuum of associated struts (orange wireframe.) (***F***) A zoomed view of the node as previously colored looking in normal to the plane of the lone G-quartet.